\documentclass[conference]{IEEEtran}
\IEEEoverridecommandlockouts
\usepackage{cite}
\usepackage{amsmath,amssymb,amsfonts}
\usepackage{algorithmic}
\usepackage{graphicx}
\usepackage{textcomp}
\usepackage{xcolor}
\usepackage[colorlinks,bookmarksopen,bookmarksnumbered,citecolor=black, linkcolor=black, urlcolor=black]{hyperref}

\usepackage{lipsum}
\usepackage{graphicx}
\usepackage{amsmath}
\usepackage{footnote}
\usepackage{mathtools}
\usepackage{mathrsfs}     % mathscr command
\usepackage{comment}
\usepackage[subrefformat=parens,labelformat=parens]{subfig}
\usepackage{bm,bbm}
\usepackage{multirow}
\usepackage{threeparttable,booktabs}
\usepackage{blkarray}
\usepackage{tikz}
\usetikzlibrary{positioning,calc,fit,decorations.pathmorphing,shapes.geometric,shapes.gates.logic.US,calc}
\usepackage{balance}
\usepackage{courier}                                       % courier font, used in \texttt
\usepackage{cleveref}                                      % smart citation
\usepackage[mathcal]{eucal}
\usepackage{algorithm}
\usepackage{filecontents}                                  % support to pgfplots
\usepackage{pgfplots}
\usepackage{pgfplotstable}
\usepgfplotslibrary{groupplots}
\pgfplotsset{compat=newest}
\usepackage[figuresright]{rotating}
\usepackage{xcolor,colortbl}                               % \rowcolor
\usepackage{diagbox}
\usepackage{makecell}
\usepackage[a4paper, total={184mm,239mm}]{geometry}
\def\BibTeX{{\rm B\kern-.05em{\sc i\kern-.025em b}\kern-.08em
    T\kern-.1667em\lower.7ex\hbox{E}\kern-.125emX}}
\begin{document}

\newcommand{\LLS}[1]{\textcolor{black}{#1}}
\newcommand{\TODO}[1]{\textcolor{red}{#1}}
\newcommand{\DYY}[1]{\textcolor{black}{#1}}

\title{Hierarchical Decoupling Capacitor Optimization for Power Delivery Network of 2.5D ICs with Co-Analysis of Frequency and Time Domains Based on Deep Reinforcement Learning
\thanks{This work was supported by Pre-research project of ministry foundation (Grant No.31513010501).(\textit{* Corresponding Authors: Leilai Shao (leilaishao@sjtu.edu.cn) and Xiaolei Zhu (xl\_zhu@zju.edu.cn)})}
\author{
\IEEEauthorblockN{Yuanyuan Duan$^{1}$, HaiYang Feng$^{1}$,  Zhiping Yu$^{2}$, Hanming Wu$^{1}$, Leilai Shao$^{3*}$, Xiaolei Zhu$^{1*}$}
    \IEEEauthorblockA{$^{1}$School of Micro-Nano Electronics, Zhejiang University, Hangzhou, China}
    \IEEEauthorblockA{$^{2}$School of Integrated Circuits, Tsinghua University, Beijing, China}
    \IEEEauthorblockA{$^{3}$School of Mechanical Engineering, Shanghai Jiao Tong University, Shanghai, China}
       % \IEEEauthorblockA{$^{*}$Corresponding Authors: Leilai Shao (leilaishao@sjtu.edu.cn) and Xiaolei Zhu (xl\_zhu@zju.edu.cn)}
    }
}
\maketitle

\begin{abstract}

2.5D integration introduces significant challenges due to increasing data rates and a large number of I/Os, necessitating advanced optimization of the power delivery networks (PDNs) both on-chip and on-interposer to mitigate the small signal noise and simultaneous switching noise (SSN). Traditional PDN optimization strategies in 2.5D systems primarily focus on reducing impedance by integrating decoupling capacitors (decaps) to lessen small signal noise. Unfortunately, relying solely on frequency-domain analysis has been proven inadequate for addressing coupled SSN, as indicated by our experimental results. In this work, we introduce a novel two-phase optimization flow using deep reinforcement learning to tackle both the on-chip small signal noise and SSN. Initially, we optimize the impedance in the frequency domain to maintain the small signal noise within acceptable limits while avoiding over-design. Subsequently, we refine the PDN in the time domain to minimize the voltage violation integral (VVI), a more accurate measure of SSN severity. To the best of our knowledge, this is the first dual-domain optimization strategy that simultaneously addresses both the small signal noise and SSN propagation through strategic decap placement in on-chip and on-interposer PDNs, offering a significant step forward in the design of robust PDNs for 2.5D integrated circuits (ICs).

\end{abstract}


\section{Introduction}

Recently, 2.5D integration has emerged as a solution to address the increasing cost of large Systems-on-Chip (SoCs) on advanced technology nodes. However, as data rates continue to increase to hundreds of gigabits per second and the number of input/outputs (I/Os) surges, maintaining the power and signal integrity poses a significant challenge for 2.5D power delivery network (PDN) design. 

\begin{figure}[t]
\centerline{\includegraphics[width=1\linewidth]{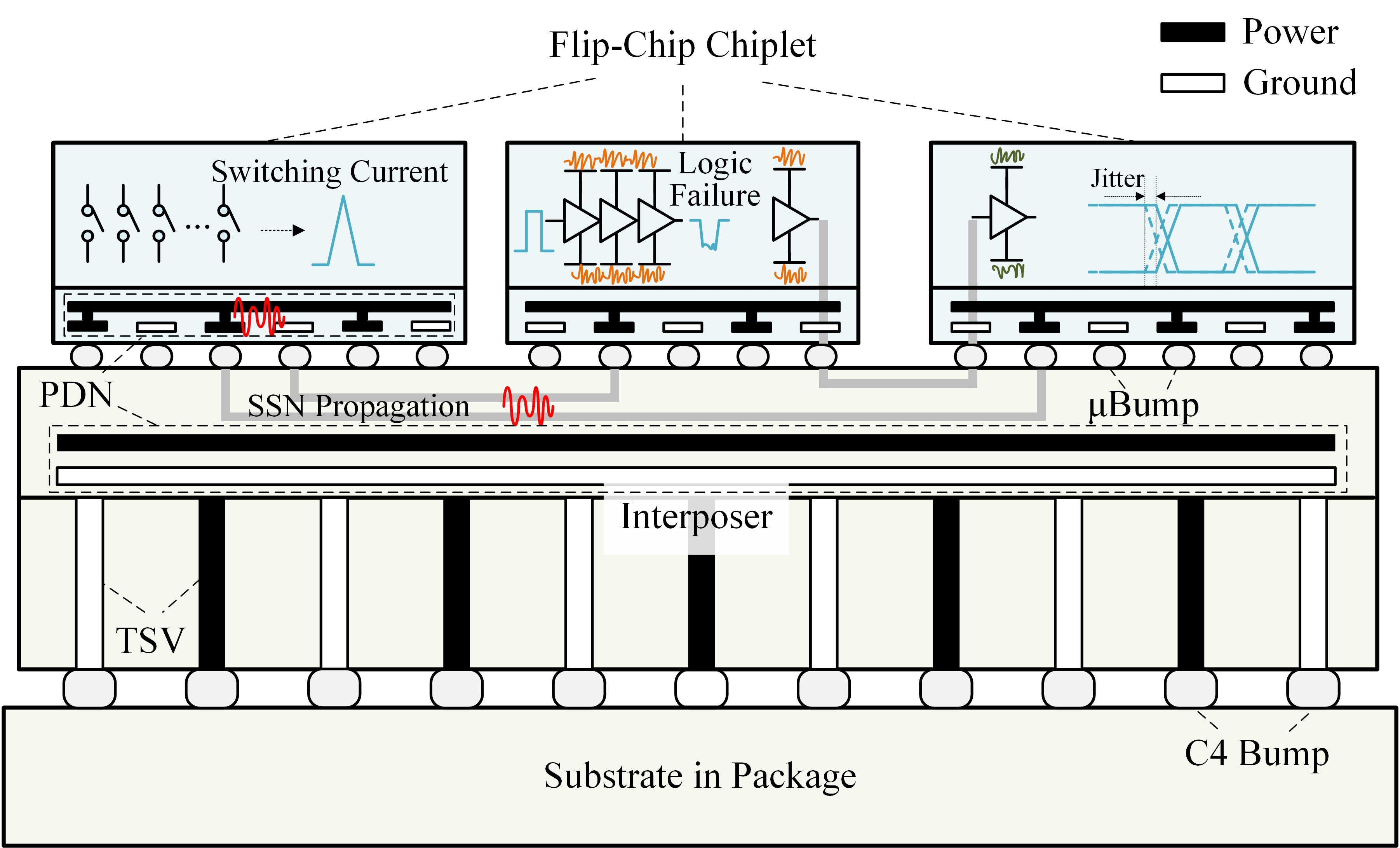}}
\caption{Cross-sectional view of 2.5D system. Large SSN generated can propagate through the hierarchical PDN and cause logic failure and jitter.}
\label{fig:2.5Dsystem}
\end{figure}

The 2.5D PDN comprises on-chip PDNs, $\mu$bumps, an on-interposer PDN, and a through-silicon via (TSV) array connecting the interposer and the package. The interposer PDN supplies power to the on-chip PDN, which in turn delivers voltages to each cell in the design. Components in electronic systems, such as voltage regulator modules (VRMs), and interconnects, introduce inductive and capacitive effects across different frequency ranges \cite{swaminathan2007power}. These effects can lead to dynamic voltage fluctuations, commonly referred to as small signal noise, which has significant implications for system performance and functionality. Furthermore, as the number of I/Os and data transmission frequencies escalate, simultaneous switching noise (SSN) becomes a critical concern, generating additional voltage fluctuations that may interfere with the operation of other chiplets. SSN, induced by the large switching currents of multiple I/Os during high-speed data transmission, can propagate through the hierarchical PDN, cause jitter \cite{kim2017static} and even logic failure \cite{hung2021power,xu2023study}, as depicted in Fig.~\ref{fig:2.5Dsystem}. Decoupling capacitors (decaps) are widely used to mitigate voltage fluctuations and help compensate for transient current demands. The hierarchical structure of the 2.5D PDN necessitates a decap strategy that optimizes the locations and capacitance of both on-chip and on-interposer decaps.

% impedance and VVI optimizaiton
PDN analysis is crucial for the design of 2.5D integrated circuits (ICs). Frequency-domain impedance often serves as a pivotal criterion for evaluating PDN reliability \cite{smith1999power,kim2013improved}. Traditional PDN optimization strategies \cite{piersanti2017decoupling,park2018reinforcement,park2020deep,park2022transformer,zhang2023decoupling} primarily focus on the impedance reduction by implementing additional decaps to alleviate the small signal noise based on the analysis of frequency domain. However, relying solely on meeting target impedance, which guarantees that voltage fluctuation remains within allowable limits, may not sufficiently consider the impact of transient responses on the overall system. The coupled SSN from adjacent chiplets' PDNs, particularly noise propagation through super high bandwidth I/Os, can lead to excessive voltage fluctuation beyond permissible levels, leading to system failure. Additionally, the high integration and miniaturization of 2.5D systems often result in the routing region occupying a significant portion of the circuit layout, constraining decap placement. Therefore, a comprehensive approach that integrates hierarchical decap placement with considerations of both the small signal noise and SSN is crucial for effectively mitigating power supply noise and ensuring the reliability of PDN designs in 2.5D systems.

% summation
In this paper, we propose a novel hierarchical decap optimization method for 2.5D systems, integrating both frequency and time domain analyses. Our approach leverages advanced deep reinforcement learning (DRL) techniques and accurately models the load current of the entire system. Source code, configurations, and detailed experimental settings are available on Anonymous GitHub: \href{https://anonymous.4open.science/r/decap_opt}{https://anonymous.4open.science/r/decap\_opt}. The key contributions of this work are summarized as follows:
\begin{itemize}
\item We present an RL-based approach for the co-optimization of on-chip and on-interposer PDNs to address both small signal noise and SSN in 2.5D ICs.
\item In the frequency domain, this approach optimizes decap placement to reduce the PDN impedance below the target impedance at probing ports, ensuring effective power delivery while avoiding unnecessary over-design.
\item In the time domain, we conduct detailed transient current simulations and introduce the voltage violation integral (VVI) as a metric. Experiments reveal that despite frequency-domain optimization, voltage violations persist. To mitigate this, we refine the PDN by strategically adding more decaps to minimize the VVI.
\item Extensive validations demonstrate that, compared to the frequency-domain optimization alone, the dual-domain optimization strategy can better mitigate small signal noise and SSN.
\end{itemize} 

The remainder of the paper is organized as follows. \Cref{sec:preliminary} presents the preliminaries. \Cref{sec:methodology} describes the details of proposed methodology. \Cref{sec:experiments} discusses the experimental results, and \Cref{sec:conclusion} draws the conclusion.
\section{Preliminaries}
\label{sec:preliminary}
\subsection{Modeling of 2.5D PDN}
\label{sec:modeling}

Modeling the 2.5D hierarchical PDN encompasses on-chip power/ground (P/G) planes, on-interposer P/G planes, TSVs, $\mu$bumps, and decaps. Each component is modeled individually and subsequently cascaded together to form the complete PDN model. The on-chip and on-interposer P/G planes are segmented into unit cells (UCs) and modeled using transmission-line (TL) theory \cite{cho2019fast,he2016extract}, where each UC is represented by unit-length resistance, inductance, conductance, and capacitance. 
The width and spacing of on-interposer P/G planes are set to 95 $\mu$m and 200 $\mu$m, while the width and spacing of on-chip P/G planes are set to 10 $\mu$m and 20 $\mu$m. 
TSVs are modeled using resistance, capacitance, and inductance \cite{kim2012modeling}, with dimensions of 100 $\mu$m in height, 20 $\mu$m in diameter, and a 200 $\mu$m pitch. The $\mu$bumps, characterized by inductance and resistance \cite{zhi2022trade}, are modeled with a height of 30 $\mu$m, a diameter of 60 $\mu$m, and a 200 $\mu$m pitch. 
For decoupling capacitors, metal-insulator-metal (MIM) capacitors are used in the interposer PDN, while metal-oxide-semiconductor (MOS) capacitors are suitable for on-chip PDNs. The smallest layout region designated for decap placement is referred to as a unit decap cell (UDC). Both chiplet and interposer UDCs are standardized to 1 mm $\times$ 1 mm to simplify design and reduce the layout complexity. The allowable capacitance for MIM capacitors on the interposer ranges from 200 pF to 2000 pF, with increments of 200 pF. Similarly, the capacitance for on-chip MOS capacitors ranges from 50 pF to 500 pF, with increments of 50 pF. Table~\ref{tab:modeling} provides a summary of modeling parameters based on the 55 nm technology node.

\begin{table}[t]
\caption{Modeling parameters of the 2.5D PDN based on 55 nm technology}
\begin{center}
\resizebox{0.8\linewidth}{!}{%
\begin{tabular}{|c|c|c|}
\hline
\textbf{Objective} 			& \textbf{Parameter} & \textbf{Value} \\ \hline
\multirow{4}{*}{Unit cell of on-chip PDN} 		& $R_{chip}$		&  		19.11 m$\Omega$\\ \cline{2-3}
 									& $L_{chip}$    		&  		8.8 pH		\\ \cline{2-3}
	 								& $G_{chip}$   		&  		2$\pi fC_{chip}tan(\delta)$		\\ \cline{2-3}
		 							& $C_{chip}$    		&  		17.7 fF		\\ \hline
\multirow{4}{*}{Unit cell of on-interposer PDN} 	& $R_{intp}$   		&  		34.2 m$\Omega$ \\ \cline{2-3}
 									& $L_{intp}$  		&  		0.63 pH		\\ \cline{2-3}
	 								& $G_{intp}$ 		&  		2$\pi fC_{intp}tan(\delta)$		\\ \cline{2-3}
		 							& $C_{intp}$ 		&  		2.79 pF		\\ \hline
\multirow{5}{*}{P/G TSV} 					& $R_{TSV}$   		&  		5.57	m$\Omega$	\\ \cline{2-3}
 									& $L_{TSV}$    		&  		30 pH			\\ \cline{2-3}
	 								& $C_{TSV}$     	&  		0.24 pF			\\ \cline{2-3}
		 							& $R_{bump}$     	&  		13.85 m$\Omega$ 	\\ \cline{2-3}
									& $L_{bump}$     	&  		2.77 pH			\\ \hline
\multirow{2}{*}{$\mu$bump} 				& $R_{\mu bump}$	&  		0.2 m$\Omega$	\\ \cline{2-3}
		 							& $L_{\mu bump}$  	&  		5.69 pH			\\ \hline										
\multirow{2}{*}{MOS capacitor} 		  & $C_{MOS}$   	& 		14.4 fF/$\mu \text{m}^2$	\\ \cline{2-3}
                                    & ESR    & 24 $\Omega$/pF\\\hline

MIM capacitor  						& $C_{MIM}$   		& 		5 fF/$\mu \text{m}^2$	\\ \hline
\end{tabular}}\label{tab:modeling}
\end{center}
\footnotesize{*$f$ is the frequency and $tan(\theta)$ is the loss tangent of dielectric.}
\end{table}

\subsection{Frequency-Domain Impedance Analysis}

To ensure a stable voltage supply for the chiplet, the impedance of PDN must remain below the target impedance value within the working frequency range. The target impedance is characterized by a flat region and a slope region \cite{smith1999power}. The target impedance in the flat region is typically defined as the ratio of the maximum allowable ripple voltage to half of the maximum transient current, $I_{max}$, derived from peak power $P_{max}$, as follows:
\begin{equation}
    Z_{target}=\frac{V_{dd} \times ripple}{I_{ref}}
    \label{eq.z_target}
\end{equation}
Here, $I_{ref}=I_{max}/2=P_{max}/2V_{dd}$ represents the typical workload and avoid over-design. Achieving low impedance at high frequencies becomes unnecessary, as it can lead to over-design and increased costs. Therefore, when the frequency exceeds the knee frequency $f_{knee}=0.35/T_r$, where $T_{r}$ is the transition time of the signal, the impedance curve increases at a rate of 20 dB/dec \cite{rao2012frequency}. In this paper, the $ripple$ and $f_{knee}$ are set as 5\% and 3.4 GHz respectively.

\subsection{Time-Domain VVI Analysis}

\begin{figure}[t]\centering
    \begin{minipage}[b]{0.7\linewidth}
        \includegraphics[width=\textwidth]{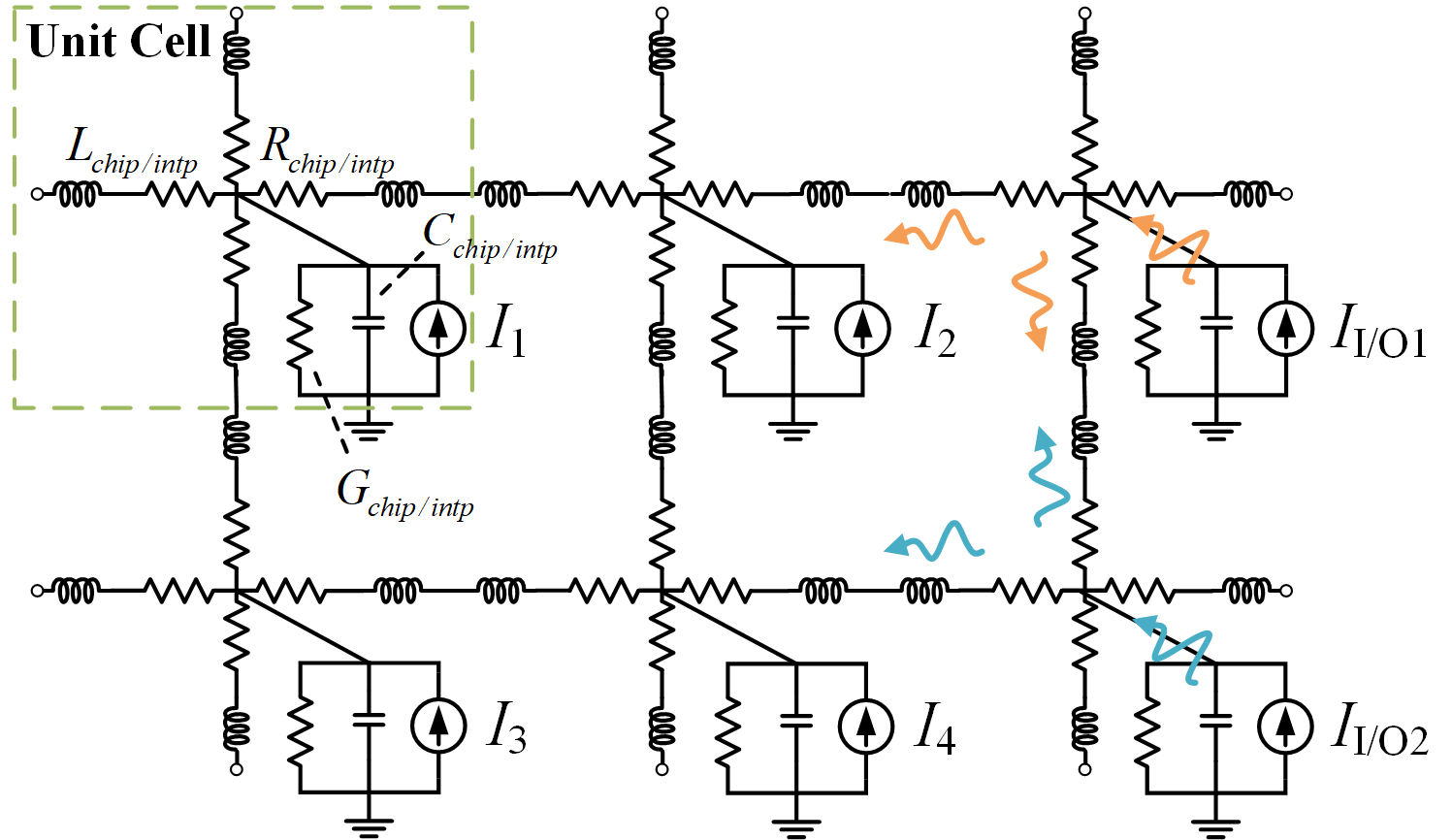}
        \centerline{(a)}
    \end{minipage}%
    \\
    \begin{minipage}[b]{0.48\linewidth}
        \includegraphics[width=\textwidth]{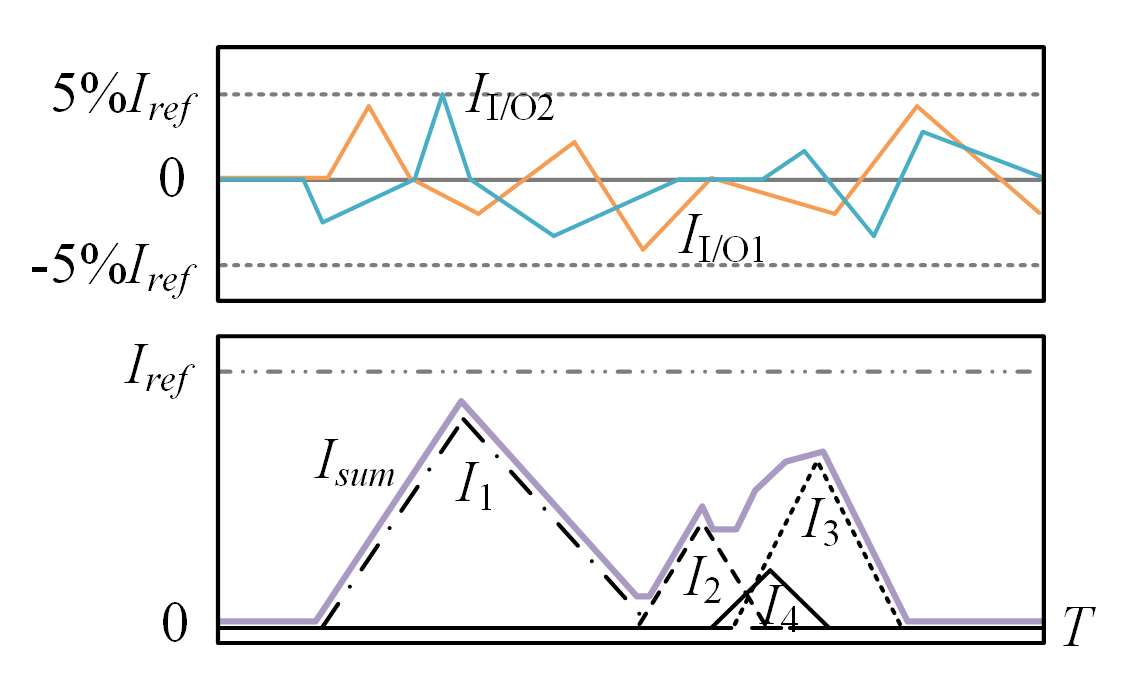}
        \centerline{(b)}
    \end{minipage}
    \begin{minipage}[b]{0.48\linewidth}
        \includegraphics[width=\textwidth]{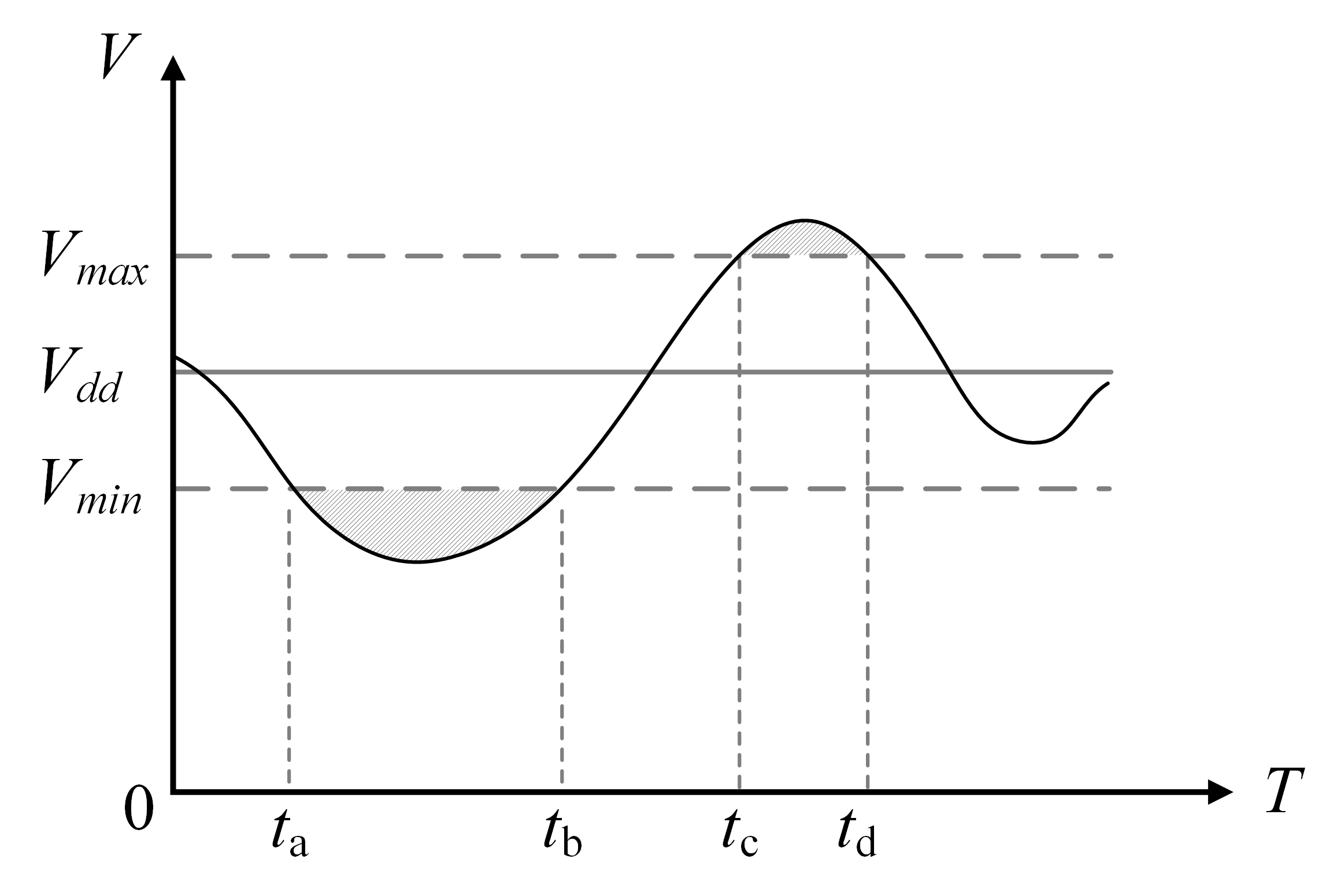}
        \centerline{(c)}
    \end{minipage}
    \caption{(a) The equivalent transmission line model with the transient currents. (b) The waveform of the internal currents and I/O currents. (c) Illustration of the voltage violation integral at a node in the $V_{dd}$ power grid.}
    \label{fig:time-domain}
\end{figure}

Frequency-domain analysis, inherently a steady-state analysis, fails to account for the effects of transient responses on circuits. In 2.5D systems, where numerous I/Os facilitate signal communication between chiplets, SSN could propagate through these I/Os, potentially disrupting the normal operation of other chiplets. SSN-induced transient current variations during operation can exceed the normal operating current, with unpredictable waveform and peak values. Consequently, accurately predicting whether voltage supply meets design requirements is challenging when relying solely on frequency-domain analysis. To ensure a robust PDN, a time-domain evaluation method is necessary.

To simulate the transient current resulting from transistor state switching \cite{su2003optimal}, we employ piecewise linear (PWL) triangular waveform currents with varying peak values and excitation times. The total current of a chiplet is modeled as the superposition of two components: the internal current ($I$) of the chiplet and the I/O current ($I_{I/O}$) fluctuations distributed at the edge of the chiplet, which can be defined as follows: 
 \begin{equation}
\text{Current Model:}
 \begin{cases}
 \begin{aligned}
&I_{sum}(t)=\sum_{i} I_{i}(t) \leq I_{ref},\\
&I_{I/Os}(t) = \lvert \sum_{i} I_{I/Oi}(t) \rvert \leq I_{ref},\\
&0\leq I_{sum}(t)+I_{I/Os}(t) \leq I_{max},\\
&\int I_{I/Oi}(t)~dt = 0.
 \end{aligned}
 \end{cases}
 \label{eq.curr_model}
\end{equation}
Here, the reference current ($I_{ref}$) is in consistency with frequency-domain analysis. To maintain overall power consumption within the $P_{max}$ limits during simulation, the sum ($I_{sum}$) of the internal currents is set below $I_{ref}$, while the I/O currents can fluctuate within a range of 5\%$I_{ref}$ with the peak value of the current summation below $I_{ref}$. Different data transmission scenarios can be modeled by specifying $I_{I/Oi}$ with varying degrees of correlation. Fig.~\ref{fig:time-domain} (a) and (b) illustrates the equivalent circuit and the waveform of the chiplet PDN under transient currents.

To conduct the dynamic power integrity analysis, we introduce the concept of voltage violation integral (VVI), which serves as a measure of the cumulative effect of voltage deviations from specified voltage fluctuations. Fig.~\ref{fig:time-domain}(c) illustrates the voltage violation at a node in the voltage supply $V_{dd}$ power grid. The VVI is calculated as the integral of the shaded area: 
\begin{equation}
V\!V\!I\!=\!\int_{0}^{T} \!\left[\max(V_{min}-V(t),0)+\max(V(t)-V_{max},0)\right]dt\label{equ:VVI}
\end{equation}
where $V_{max}$ and $V_{min}$ represent the maximum and minimum allowable voltages, set to 105\% and 95\% of the power supply voltage, respectively.
As the duration of time during which the voltage exceeds the maximum allowable voltage fluctuation range increases, the likelihood of circuit errors also rises. Considering both the magnitude and duration of voltage deviations, the VVI provides a comprehensive measure of PDN performance, especially in dynamic operating conditions characterized by transient events. 

% Compared to the methods solely considering the worst-case voltage drop over time \cite{kim2021chiplet}, the VVI analysis can assess the overall severity of dynamic voltage violations and evaluate the effectiveness of PDN optimization strategies in mitigating SSNs. Thus, minimizing the VVI could ensure the reliability and performance of electronic systems, particularly in high-speed and high-density applications where voltage fluctuations can lead to system failure or performance degradation.

\section{Methodology}
\label{sec:methodology}

The proposed RL-based method for 2.5D PDN decap optimization aims to minimize total decap capacitance while meeting target impedance and reducing the VVI. This optimization problem can be formulated as a Markov decision process (MDP), defined by the state $S$, action $A$, and reward $R$. The detailed algorithm process is described in \Cref{sec:process}. Definitions of parameters for impedance and VVI optimization are provided in \Cref{sec:impedance matrix} and \Cref{sec:VVI matrix}, respectively. The deep neural network (DNN) structure employed in the proposed RL algorithm is discussed in \Cref{sec:structure}.

%Fig.~\ref{fig:method} illustrates the RL-based approach for optimizing the PDN of 2.5D ICs.
% \begin{figure}[t]
% \centerline{\includegraphics[width=1\linewidth]{fig/method.jpg}}
% \caption{The overall concept of the RL-based method for 2.5D PDN decap optimization.}
% \label{fig:method}
% \end{figure}

\subsection{Algorithm Process for Decap Optimization}
\label{sec:process}
The general algorithm process is described as follows.
\begin{itemize}
\item[(1)]
\textbf{Early-Stage Floorplanning}: Determine the early-stage floorplanning, yielding complete designs for the interposer and chiplets, including placement, routing, and PDNs.
\item[(2)]
\textbf{Hierarchical PDN Modeling}: Generate the 2.5D hierarchical PDN models in RLGC format, as discussed in \Cref{sec:modeling}.
\item[(3)]
\textbf{Impedance Analysis}: Perform impedance analysis utilizing the circuit simulator NGSPICE. Optimize the locations and capacitance of on-chip and on-interposer decaps to meet the target impedance requirements.
\item[(4)]
\textbf{VVI Optimization}: Simulate transient currents with NGSPICE. Then, refine the on-chip decap placement to minimize VVI in the time domain.
\end{itemize}

\subsection{Matrix Definition Based on Impedance Analysis}
\label{sec:impedance matrix}
For a given hierarchical PDN, impedance analysis considers several factors: the chiplet layout, interposer space (considering non-capacitor zones), and the distribution of MIM capacitors on the interposer PDN and MOS capacitors on the chiplet PDN. This information is encoded as the state $S_I$ into four 2D matrices: the Interposer Space Matrix, the Chiplet Space Matrix, the MIM Distribution Matrix, and the MOS Distribution Matrix. The dimensions of these matrices are determined by the number of UDCs. The Space Matrix delineates feasible decap locations on the interposer or chiplet layer and is a binary matrix, where '1' indicates feasible positions and '0' denotes non-feasible locations. The Distribution Matrix represents the normalized capacitance values of decaps, where '0' indicates the absence of a unit decap and '1' represents the presence of a unit decap with the maximum allowable capacitance.

The action is defined as the change in capacitance of unit decaps at each timestep. There are ten distinct, incrementally increasing capacitance levels for both MIM and MOS capacitors. The action space $A_I$ encompasses all potential combinations of these changes across all unit decaps of the on-chip and on-interposer PDNs, expressed as:
\begin{equation}
\{-c_{MOS}/c_{MIM},0,+c_{MOS}/c_{MIM}\}^{N_{chip}+N_{intp}}\label{equ:space}
\end{equation}
Here, $N_{chip}$ and $N_{intp}$ denote the number of available UDCs of on-chip and on-interposer PDNs, respectively. Each unit decap can either increase, decrease, or maintain its capacitance by a step size defined by the ratio of $c_{MOS}$ and $c_{MIM}$.

To monitor the impedance variations, probing ports $\bm{P}$ are strategically placed across different chiplets. The goal is to ensure that the impedance measured at all ports meets the target impedance across the frequency range of interest, while minimizing the manufacturing cost and the leakage current induced by excessive decaps. The RL agent is trained to increase capacitance when impedance exceeds the target and to use minimal capacitance when the impedance meets the target. Therefore, the reward function $R_I$ is defined as:
\begin{equation}
R_I\!=\!
\begin{cases}
\begin{aligned}
&-\sum_f\mathop{\max}_{\bm{P}}(Z(f)-Z_{target}(f)), \text{ if $Z$ is not satisfied}\\
&\alpha(1-\frac{\sum C_{mos}}{\sum C_{chip_m}})+\beta(1-\frac{\sum C_{mim}}{\sum C_{intp_m}}), \text{ otherwise}
\end{aligned}
\end{cases}\label{equ:rewardf}
\end{equation}
where $Z-Z_{target}$ is the difference between the actual and target impedance observed at $\bm{P}$ across frequencies $f$, ranging from 0.1 to 20 GHz, with 100 points sampled per decade. $\sum C_{mos}$ and $\sum C_{chip_m}$ denote the total placed and maximum allowable capacitance for MOS capacitors, while $\sum C_{mim}$ and $\sum C_{intp_m}$ represent the same for MIM capacitors. The weights $\alpha$ and $\beta$, which sum to 1, are both set to 0.5 in this paper.

\subsection{Matrix Definition Based on VVI Analysis}
\label{sec:VVI matrix}
Minimizing VVI is crucial for maintaining stable voltage levels and mitigating voltage violation effects, particularly in high-performance electronic systems. The RL agent is trained to optimize the decap configuration of the on-chip PDNs to achieve lower VVI values, building upon prior impedance optimization. To accomplish this, the VVIs of all on-chip PDN nodes are monitored for optimization. In addition to the PDN information mentioned in \Cref{sec:impedance matrix}, the VVI information is also included in the input state matrices $S_V$ as a 2D matrix.

The action space $A_V$ can be adjusted by selecting which available UDCs—either on-chip or on-interposer—will place additional decaps. In subsequent experiments, which primarily focus on on-chip decaps, the action space is defined as follows:
\begin{equation}
\{ -c_{MOS},0,+c_{MOS}\}^{N_{chip}}\label{equ:VVIspace}
\end{equation}
% The action $A_V$ can follow the same definition as in impedance analysis. Notably, there is considerable flexibility in selecting the position of the decap (on-chip or on-interposer) to be placed, provided the UDC is available.
% Due to their proximity to noise sources, on-chip decaps are more effective in mitigating SSN. Incorporating this locality insight, only on-chip decaps are considered in subsequent experiments, which could significantly improve convergence speed without compromising the quality of the final decap distribution. Therefore, the action space $A_V$ is defined as:
% \begin{equation}
% \{ -c_{MOS},0,+c_{MOS}\}^{N_{chip}}\label{equ:VVIspace}
% \end{equation}

To further alleviate SSN, the optimization objective is to minimize the VVIs across all nodes in the on-chip PDNs, which necessitates refining the reward function. To emphasize the improvement between the initial and optimized VVIs, the reward function $R_V$ is defined as:
\begin{equation}
R_V=
\begin{cases}
\begin{aligned}
&1-\frac{\sum V}{\sum V_{init}}, \text{ if $\frac{\sum V}{\sum V_{init}}>\gamma$}\\
&1-\gamma+(1-\frac{\sum C_{mos}}{\sum C_{chip_m}}), \text{ otherwise}
\end{aligned}
\end{cases}
\label{equ:rewardt}
\end{equation}
where $V_{init}$ and $V$ represent the VVI at a node before and after optimization, respectively. $\gamma$ represents the VVI tolerance that can be adjusted by designers to meet specific requirements. This reward function guides the agent to add the minimum amount of MOS capacitance necessary to meet the $\gamma$.

\subsection{Architecture and RL Algorithm}
\label{sec:structure}

\begin{figure}[t]
\centerline{\includegraphics[width=0.8\linewidth]{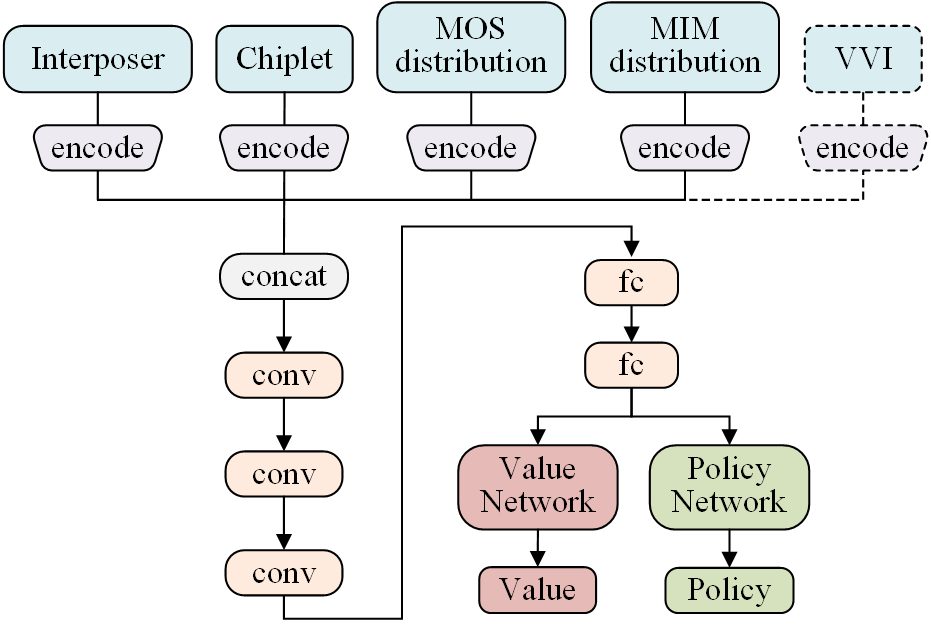}}
\caption{Feature embedding and DNN structure for the proposed RL algorithm, where VVI is only used in the time-domain optimization.}
\label{fig:network}
\end{figure}

The architecture of the proposed method, as illustrated in Fig.~\ref{fig:network}, consists of two networks: a policy network and a value network. The policy network selects actions by generating a probability distribution over available actions based on the current state, while the value network estimates the expected cumulative reward from a given state according to the current policy, providing essential feedback to the policy network. Both networks are implemented using an DNN structure, where all state information is initially encoded into matrices and then concatenated into a single matrix for input. The two networks share the same feature extraction layers, which comprise several convolutional layers, followed by fully connected layers (for detailed information, please refer to the Anonymous GitHub: \href{https://anonymous.4open.science/r/decap_opt}{https://anonymous.4open.science/r/decap\_opt}). The extracted features then pass through the policy network to generate a probability distribution, while the value network produces a value representing the quality of the policy.

We employ the proximal policy optimization (PPO) \cite{schulman2017proximal} algorithm to train the policy and value networks. The objective functions are formulated as:
\begin{equation}
L_{\text{policy}}(\theta)\!=\!\hat{\mathbb{E}} \!\left[\text{min}\left(r_t(\theta)\hat{A}_t, \text{clip}(r_t(\theta),1-\epsilon,1+\epsilon\right)\hat{A}_t\right]
\label{equ:policy}
\end{equation}
\begin{equation}
L_{\text{value}}(\phi) = \hat{\mathbb{E}} \left[ \left( R_t - V_{\phi}(s_t) \right)^2 \right]
\label{equ:value}
\end{equation}
where $r_t(\theta)=\pi(a_t|s_t)/\pi_{old}(a_t|s_t)$ denotes the ratio of the new policy and the old policy. $\hat{A}_t=R_t-V_{\phi}(s_t)$ is the advantage function, where $R_t$ is the cumulative reward from time $t$, and $V_{\phi}(s_t)$ is the value function that estimates the return for state $s_t$. Based on these loss functions, the policy and value networks are updated with the gradient descent algorithm.

%In this paper, $c_1$ and $c_2$ are both set as 0.5. The Adam optimizer is utilized for training, and the learning rate is established as $10^{-4}$. 

\section{Experiments}
\label{sec:experiments}
\subsection{Benchmarks}

\begin{figure}[t]
\centerline{\includegraphics[width=0.6\linewidth]{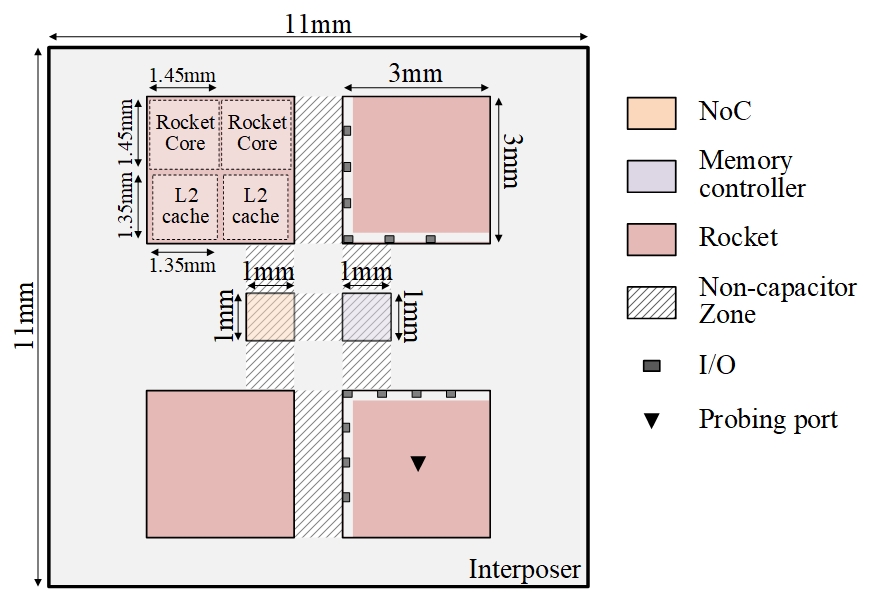}}
\caption{Rocket-64 with the non-capacitor zone. \LLS{I/Os are evenly distributed at the inner two edges and four probing ports are selected at the center of each Rocket chiplet.}}
\label{fig:rocket}
\end{figure}

To validate the proposed method, five test cases with different PDN configurations, including ROCKET-64 \cite{kim2019architecture} and four synthetic cases, are employed. The ROCKET-64 configuration includes six chiplets: a Network-on-Chip (NoC), a memory controller, and four merged Rockets, each consisting of two Rocket cores and two L2 Cache units, as illustrated in Fig.~\ref{fig:rocket}. The on-interposer PDN consists of an 11$\times$11 grid, while each on-chip PDN consists of a 3$\times$3 UDC grid.

\subsection{Frequency-Domain Impedance Optimization}

\begin{table}[t]
\caption{Comparisons of methods in the frequency domain}
\begin{center}
\resizebox{1\linewidth}{!}{%
\begin{tabular}{|c|c|c|c|c|c|c|c|c|c|}
\hline 
\multirow{3}{*}{\diagbox[height=3.2em]{\textbf{Result}}{\textbf{Method}}}
& \multicolumn{3}{c|}{\textbf{Proposed Method}}
& \multicolumn{3}{c|}{\textbf{DA}} 
& \multicolumn{3}{c|}{\textbf{GA}} \\ \cline{2-10}
&\multirow{2}{*}{\textbf{Reward}}&\textbf{MIM}&\textbf{MOS} &\multirow{2}{*}{\textbf{Reward}}&\textbf{MIM}&\textbf{MOS}
&\multirow{2}{*}{\textbf{Reward}}&\textbf{MIM}&\textbf{MOS} \\
& &(nF)&(nF)& &(nF)&(nF)& &(nF)&(nF) \\\hline

\textbf{ROCKET-64}  & 0.874 & 27.0  & 2.3  & 0.750  & 80.2  & 2.3   & 0.658  & 82.6  & 5.4 \\\hline
\textbf{Case1}      & 0.823 & 40.2  & 4.7  & 0.672  & 93.4  & 6.8   & 0.664  & 91.6  & 7.3 \\\hline
\textbf{Case2}      & 0.893 & 27.8  & 3.2  & 0.652  & 98.4  & 9.3   & 0.683  & 79.6  & 9.7 \\\hline
\textbf{Case3}      & 0.779 & 56.2  & 5.3  & 0.696  & 97.2  & 5.2   & 0.662  & 104.4 & 6.1 \\\hline 
\textbf{Case4}      & 0.761 & 60.0  & 5.8  & 0.697  & 83.8  & 6.5   & 0.711  & 80.2 &  6.2 \\\hline\hline
\textbf{Training Time} &\multicolumn{3}{c|}{5 hours} &\multicolumn{3}{c|}{10 hours}  &\multicolumn{3}{c|}{10 hours} \\ \hline
\textbf{Improvement} &\multicolumn{3}{c|}{19.41\% / 22.44\%} &\multicolumn{3}{c|}{-}  &\multicolumn{3}{c|}{-} \\ \hline
\end{tabular}}\label{tab:impedance}
\end{center}
\end{table}

To evaluate the performance of the proposed RL-based method in the frequency domain, we compare it with the dual annealing (DA) algorithm and the genetic algorithm (GA). The cost functions for DA and GA are defined as the inverse of the reward function used in the RL method. Table~\ref{tab:impedance} summarizes the comparison of the optimal performance across test cases. The reward achieved by the proposed method shows improvements of 19.41\% and 22.44\% over the solutions obtained by DA and GA, respectively. Furthermore, the proposed method required lower capacitance values and approximately 5 hours of training time, compared to around 10 hours for DA and GA. The RL-based method effectively addresses large-scale, complex optimization problems, achieving optimal performance, whereas GA and DA tend to fall into local optima and require significantly longer time to converge to the optimal solution.

\begin{figure}[t]
\centering
	\begin{minipage}{0.32\linewidth}
		\centering
		\includegraphics[width=1\linewidth]{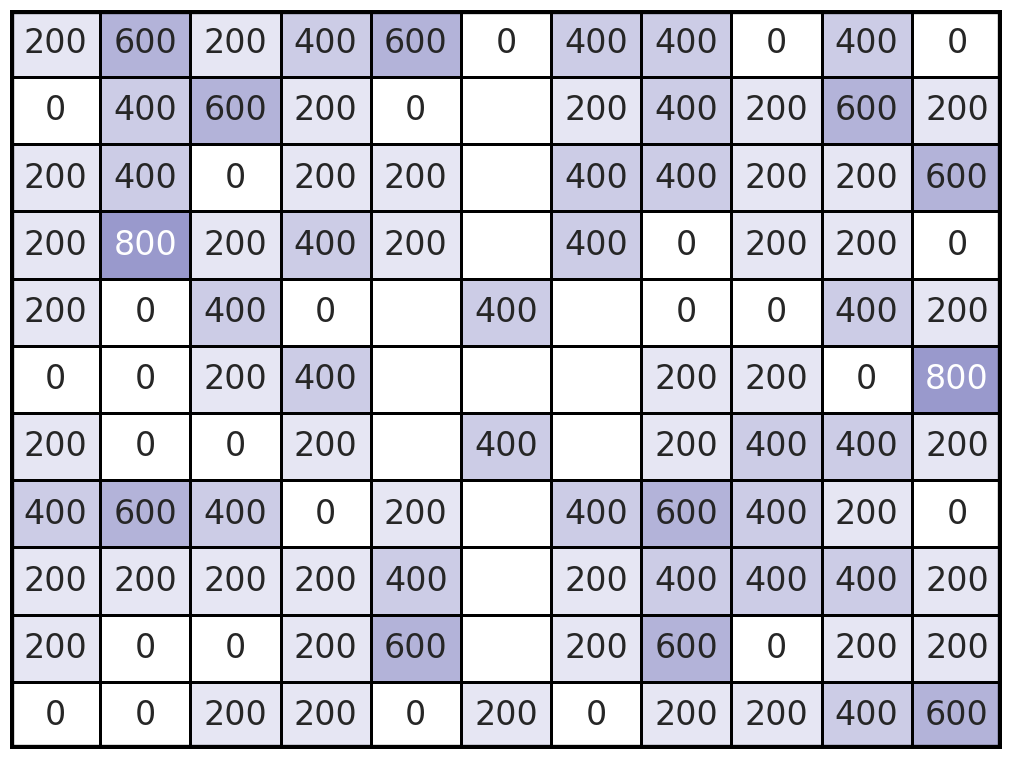}
		 \centerline{(a) RL}
		\label{fig9a}
	\end{minipage} \hfill
	\begin{minipage}{0.32\linewidth}
		\centering
		\includegraphics[width=1\linewidth]{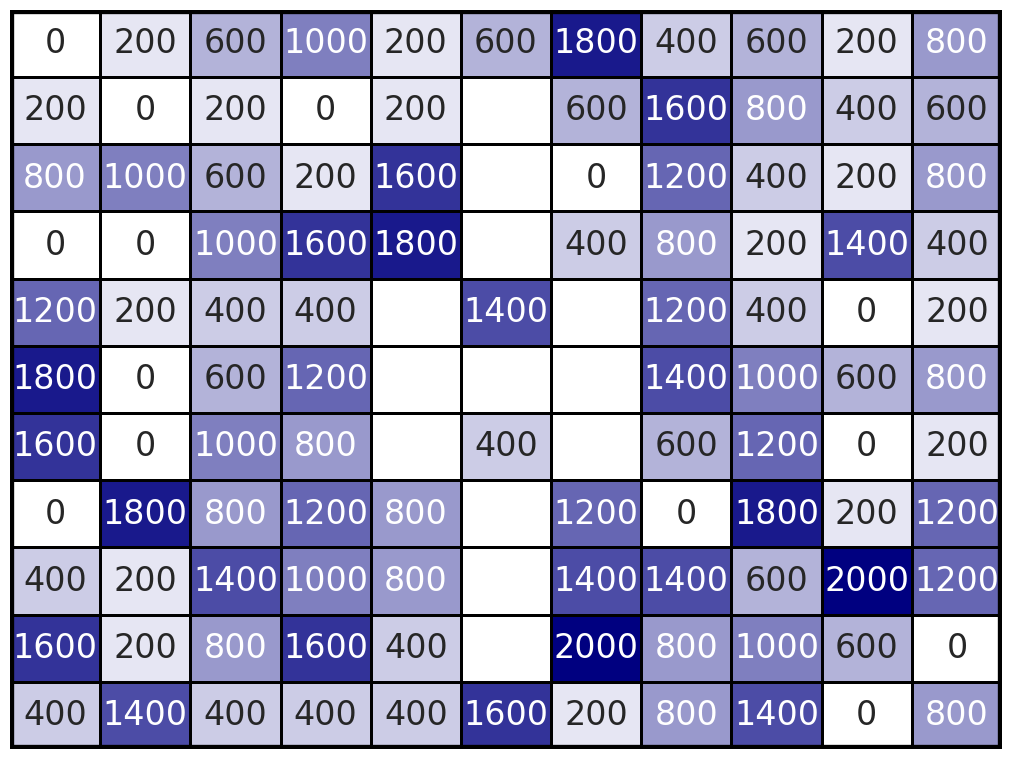}
		  \centerline{(b) DA}
		\label{fig9b}
	\end{minipage} \hfill
	%\qquad
	\begin{minipage}{0.32\linewidth}
		\centering
		\includegraphics[width=1\linewidth]{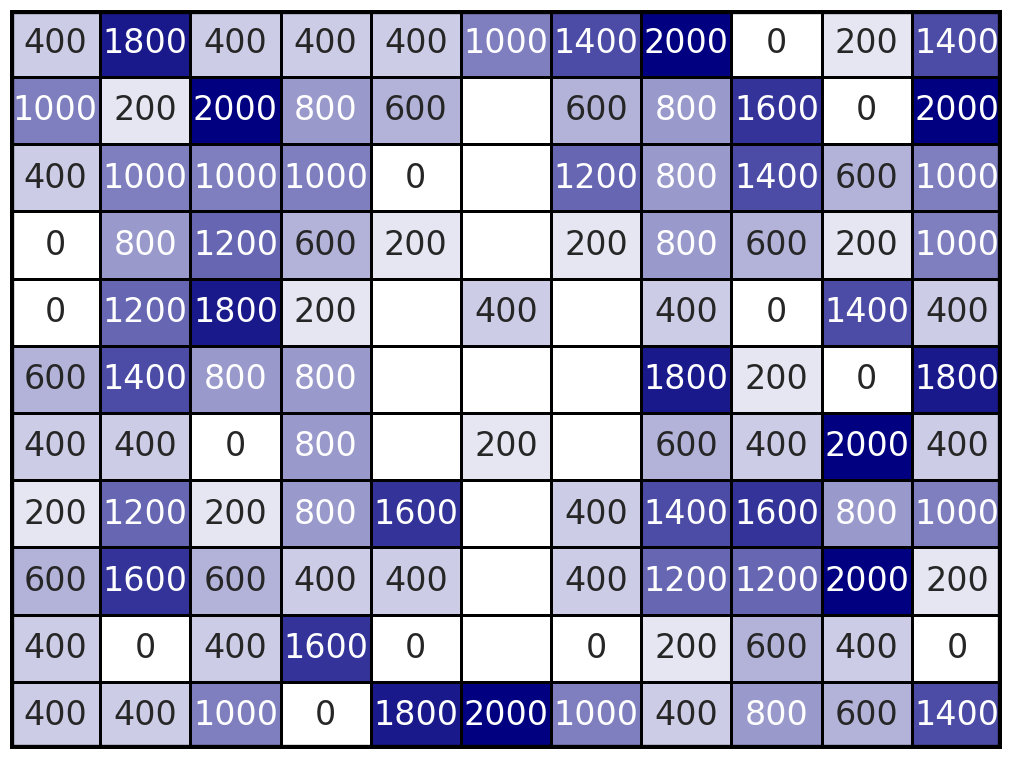}
		  \centerline{(c) GA}
		\label{fig9c}
	\end{minipage} \\
	\begin{minipage}{0.32\linewidth}
		\centering
		\includegraphics[width=1\linewidth]{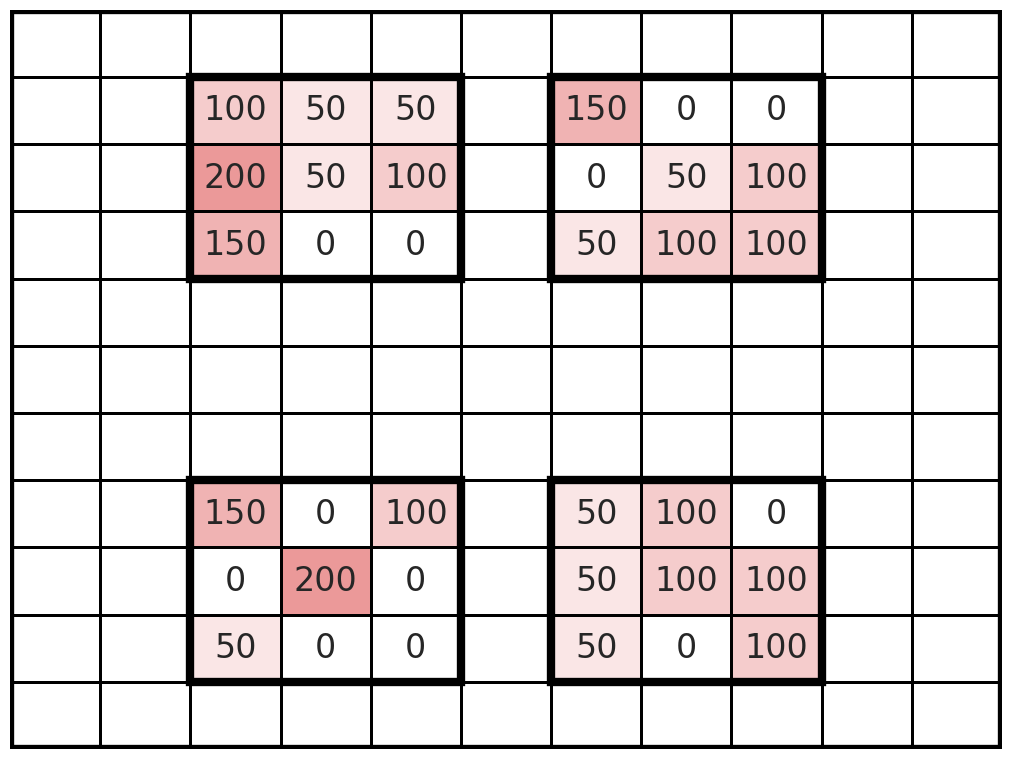}
		  \centerline{(d) RL}
		\label{fig9d}
	\end{minipage} \hfill
 	\begin{minipage}{0.32\linewidth}
		\centering
		\includegraphics[width=1\linewidth]{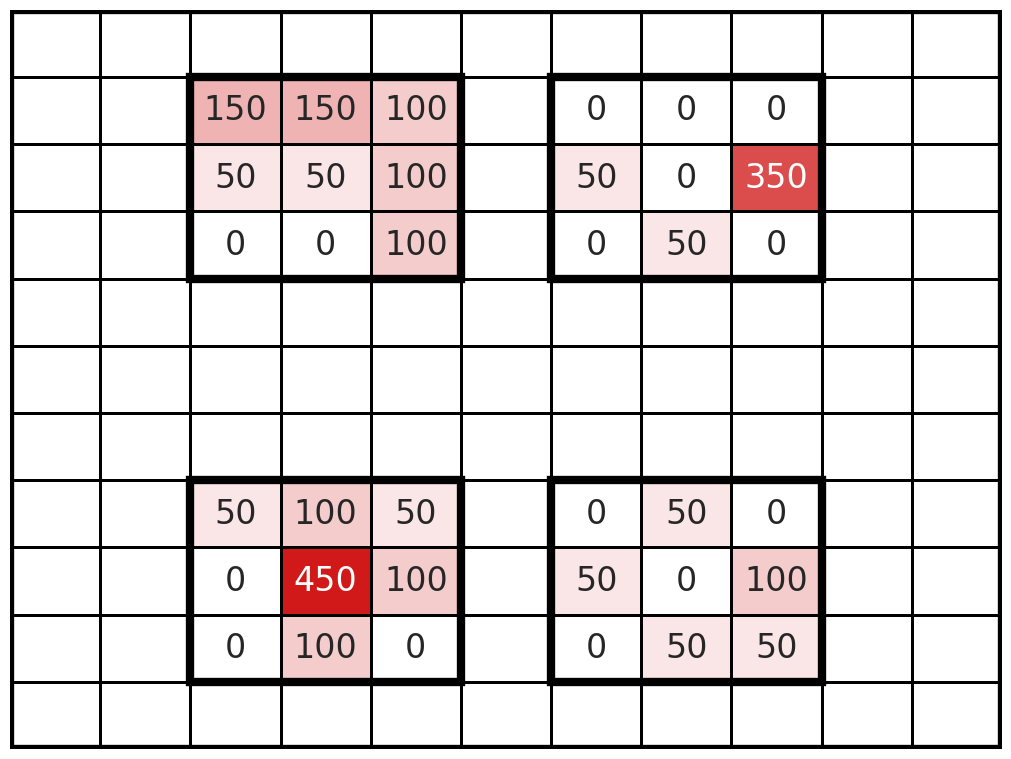}
		  \centerline{(e) DA}
		\label{fig9d}
	\end{minipage} \hfill
 	\begin{minipage}{0.32\linewidth}
		\centering
		\includegraphics[width=1\linewidth]{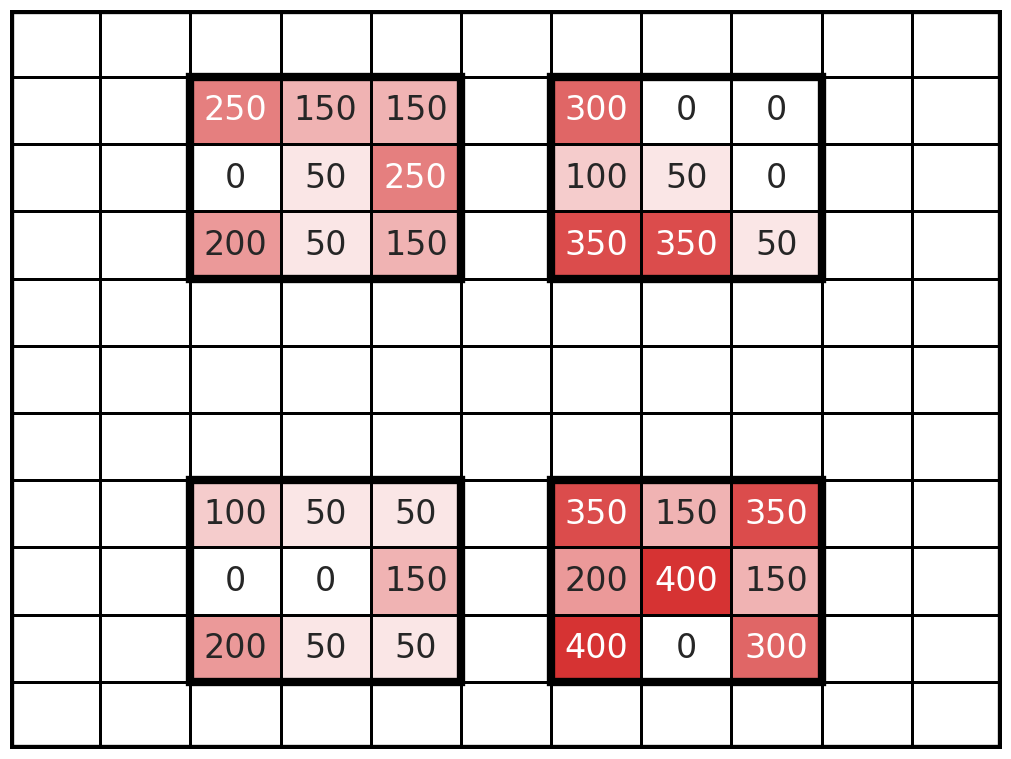}
		  \centerline{(f) GA}
		\label{fig9d}
	\end{minipage}
\caption{The decap distribution of ROCKET-64 after impedance optimization: (a)(b)(c)interposer layer; (d)(e)(f)chiplet layer. Each grid value indicates capacitance in pF, with blank grids representing non-capacitor zones.}
\label{fig:distribution}
\end{figure}

\begin{figure}[t]
\centerline{\includegraphics[width=0.5\linewidth]{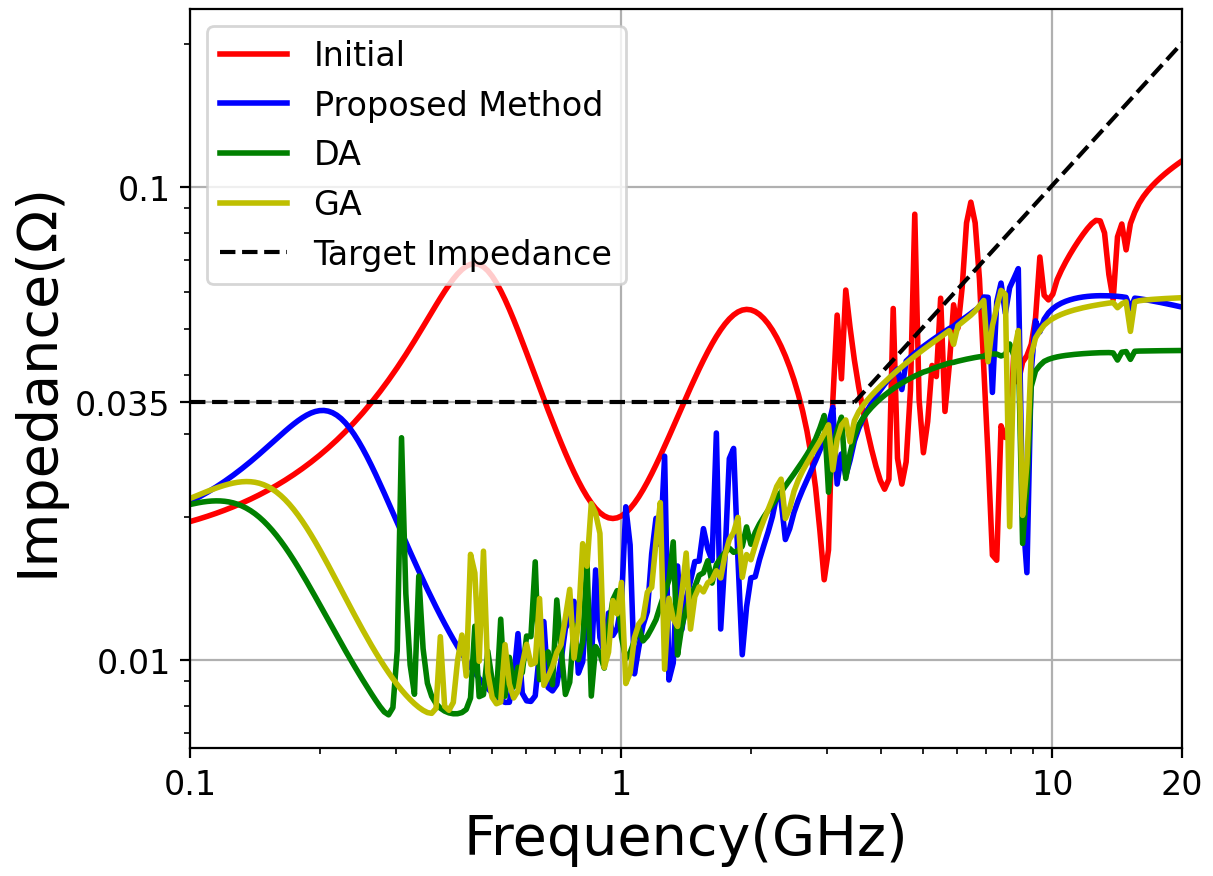}}
\caption{The impedance curves at a probing port of the initial and optimized PDN using the proposed method, DA, and GA.}
\label{fig:port impedance}
\end{figure}

The optimal decap distribution for ROCKET-64 is shown in Fig.~\ref{fig:distribution}. Compared to DA and GA, the proposed method required less layout space and lower capacitance. Fig.~\ref{fig:port impedance} shows the impedance curves at one probing port. The dashed line represents the target impedance, with a flat region of 35 m$\Omega$ and a frequency-dependent increase rate of 20 dB/dec beyond 3.4 GHz. While all methods achieved the desired solutions, the proposed method optimized the PDN without over-design, achieving the target impedance across the frequency range of 0.1 to 20 GHz with reduced capacitance, layout space, and design time.

\subsection{Time-Domain VVI Optimization}

\subsubsection{Impact of Simultaneous Switching}
To investigate the voltage violations after frequency-domain optimization, we conducted a time-domain experiment on the ROCKET-64 design. Internal current sources were generated based on the PWL current model and the $I_{ref}$ constraint, as described in Equation~\ref{eq.curr_model}, and attached to each UC of the PDN to simulate the normal operating conditions of the Rocket chiplet. Besides internal current sources, 13 lumped current sources were evenly distributed along the inner edges of each Rocket chiplet to model current fluctuations during high-speed communications. The correlation coefficient between I/O current sources was used to represent different data transmission patterns in high-speed I/Os, where higher correlation indicates more simultaneous switching of TX/RX circuits.

\begin{figure}[t]
\centering
	\begin{minipage}{0.46\linewidth}
		\centering
		\includegraphics[width=1\linewidth]{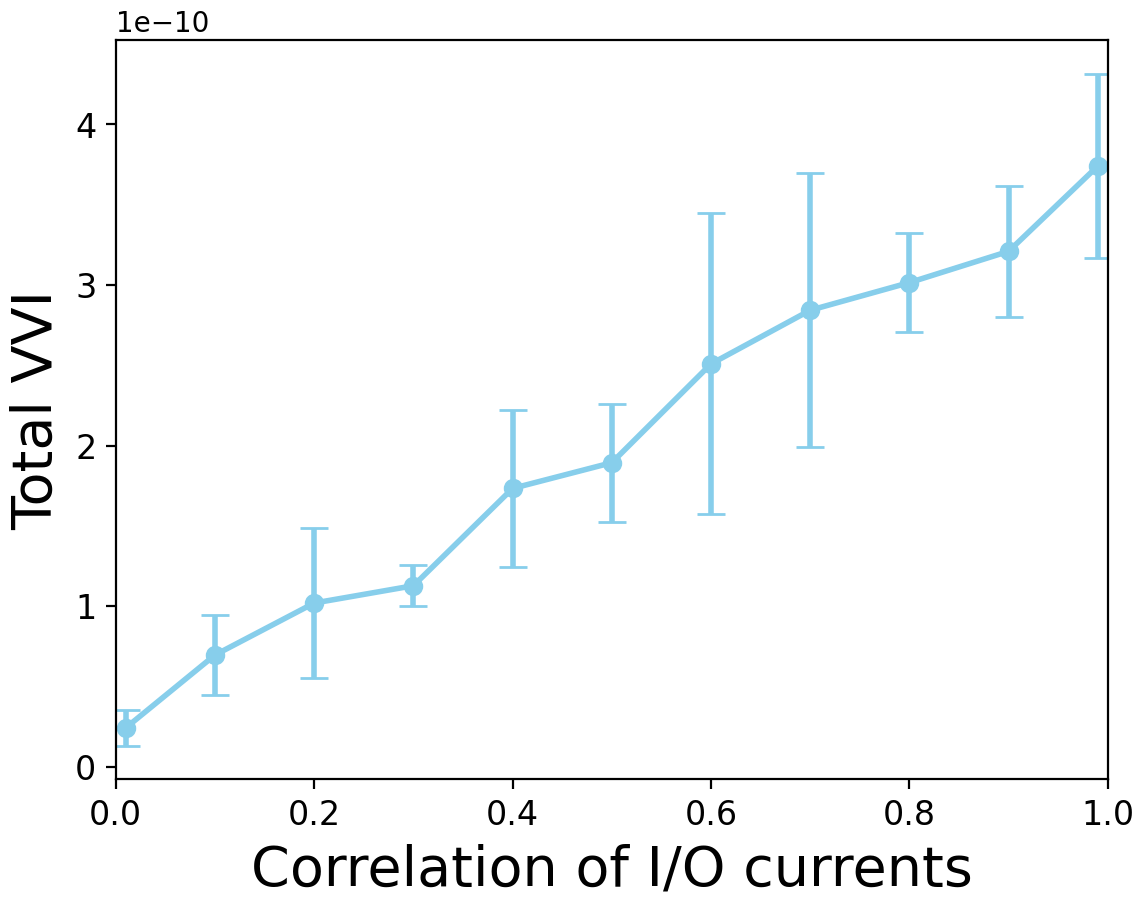}
		 \centerline{(a)}
	\end{minipage} 
	\begin{minipage}{0.49\linewidth}
		\centering
		\includegraphics[width=1\linewidth]{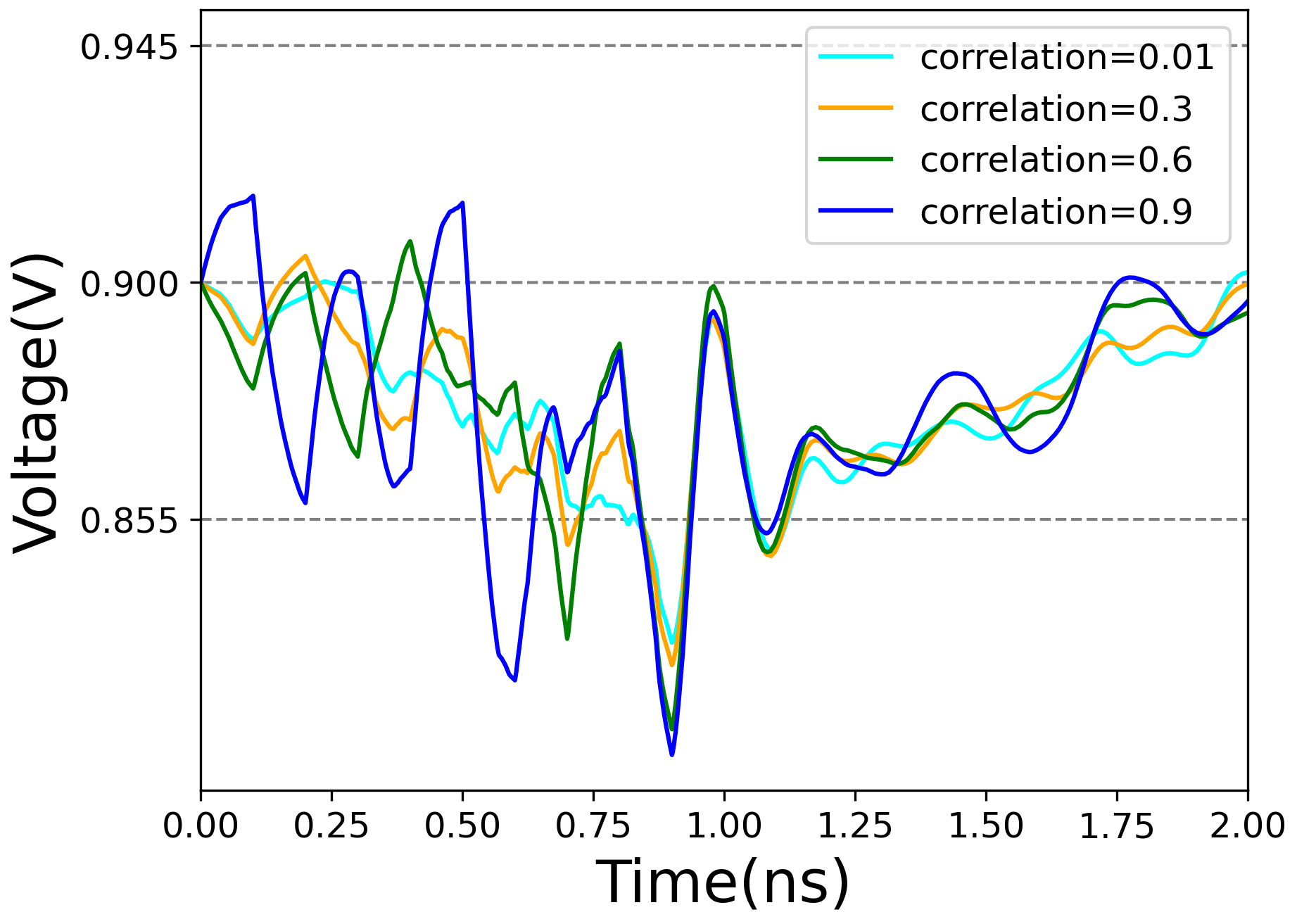}
		  \centerline{(b)}
	\end{minipage}
 \caption{The effect of simultaneous switching: (a) Total VVI variations under different I/O current correlations; (b) Voltage fluctuations of a typical node under different correlations between I/O currents.}
\label{fig:IOcurrent}
\end{figure}

We conducted extensive investigations with 50 generated current profiles for correlation coefficients ranging from 0 to 1.0 in the post-frequency-optimized PDN, as summarized in Fig.~\ref{fig:IOcurrent}(a). The total VVI of all nodes within a 2 ns interval remained above zero, regardless of the correlation level, indicating that voltage violations persist even when the target impedance is met across the entire frequency range. As the correlation among I/O currents increases, the total VVI also escalates. Monitoring a typical node in the on-chip PDN revealed that, under any level of simultaneous switching, the voltage variation exceeded the 5\% voltage ripple limit of $V_{dd}$, as depicted in Fig.~\ref{fig:IOcurrent}(b). Additionally, as simultaneous switching increased, voltage violations became more severe, demonstrating the significant impact of SSN on PDN power stability. Consequently, time-domain optimization is essential to ensure the performance of the 2.5D PDN.

\subsubsection{Results of VVI Optimization}

\begin{table}[t]
\caption{Comparisons before and after time-domain optimization on ROCKET-64}
\begin{center}
\resizebox{0.9\linewidth}{!}{%
\begin{tabular}{|c|c|c|c|}
\hline
&\begin{tabular}[c]{@{}c@{}}\textbf{Total MOS}\\ \textbf{Capacitance}\end{tabular}
&\begin{tabular}[c]{@{}c@{}}\textbf{Total}\\ \textbf{VVI}\end{tabular} 
&\begin{tabular}[c]{@{}c@{}}\textbf{Number of}\\ \textbf{Violation Nodes}\end{tabular} \\ \hline
\textbf{Before} &2.3 nF   &$3.835\times 10^{-10}$&196 \\ \hline
$\gamma$=0.50   &5.4 nF   &$1.884\times 10^{-10}$&196 \\ \hline
$\gamma$=0.20	&8.8 nF   &$7.046\times 10^{-11}$&196 \\ \hline
$\gamma$=0.10   &10.4 nF  &$3.757\times 10^{-11}$&173 \\ \hline
$\gamma$=0.05	&11.5 nF  &$1.897\times 10^{-11}$&153 \\ \hline
$\gamma$=0.02	&12.9 nF  &$6.244\times 10^{-12}$&98\\ \hline
$\gamma$=0  	&17.8 nF  &0        &       0 \\ \hline
\end{tabular}}
\label{tab:time-domain}
\end{center}
\end{table}

\begin{figure}[t]
\centering
	\begin{minipage}{0.49\linewidth}
		\centering
		\includegraphics[width=1\linewidth]{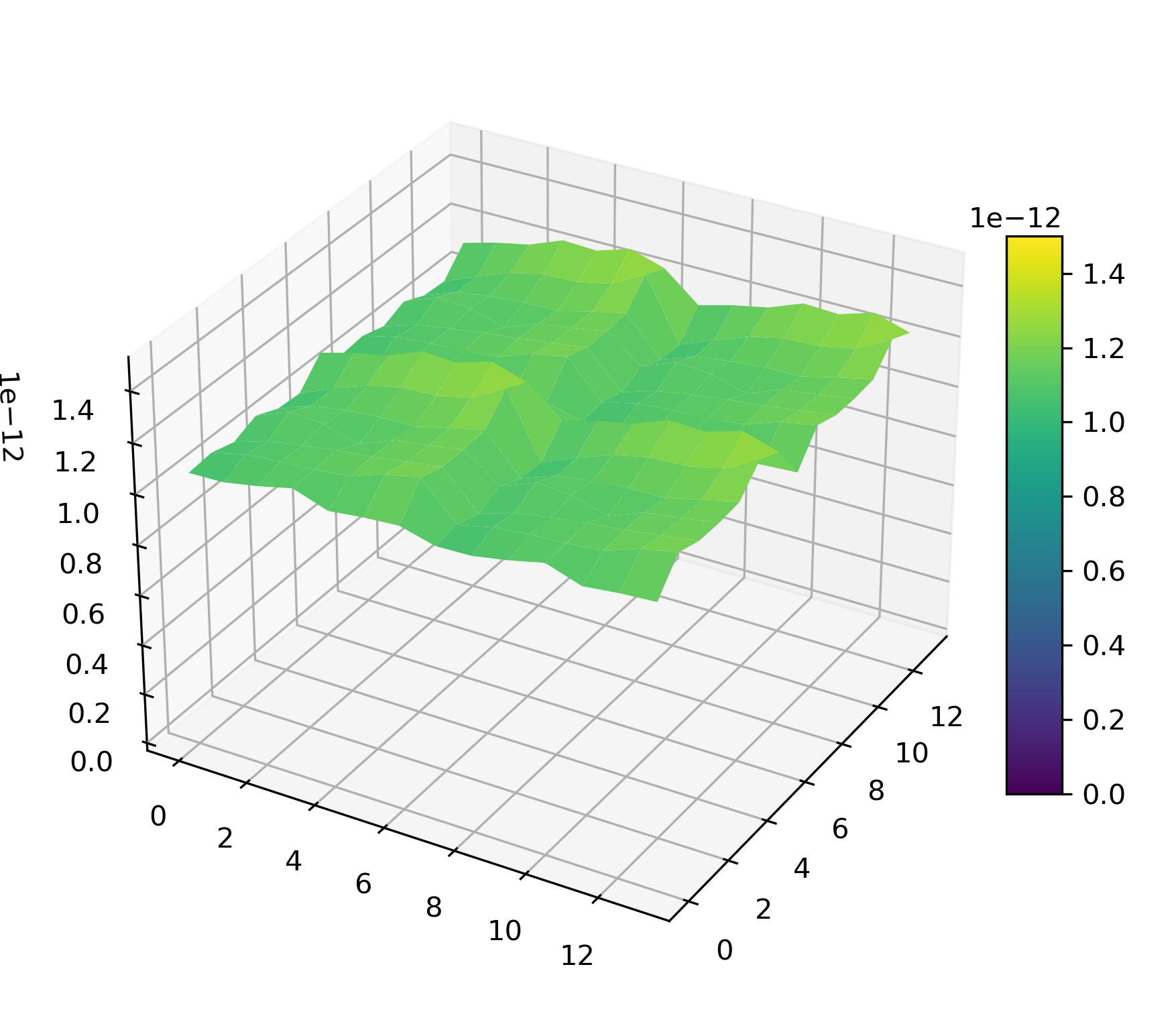}
		 \centerline{(a)Before}
	\end{minipage} 
	\begin{minipage}{0.49\linewidth}
		\centering
		\includegraphics[width=1\linewidth]{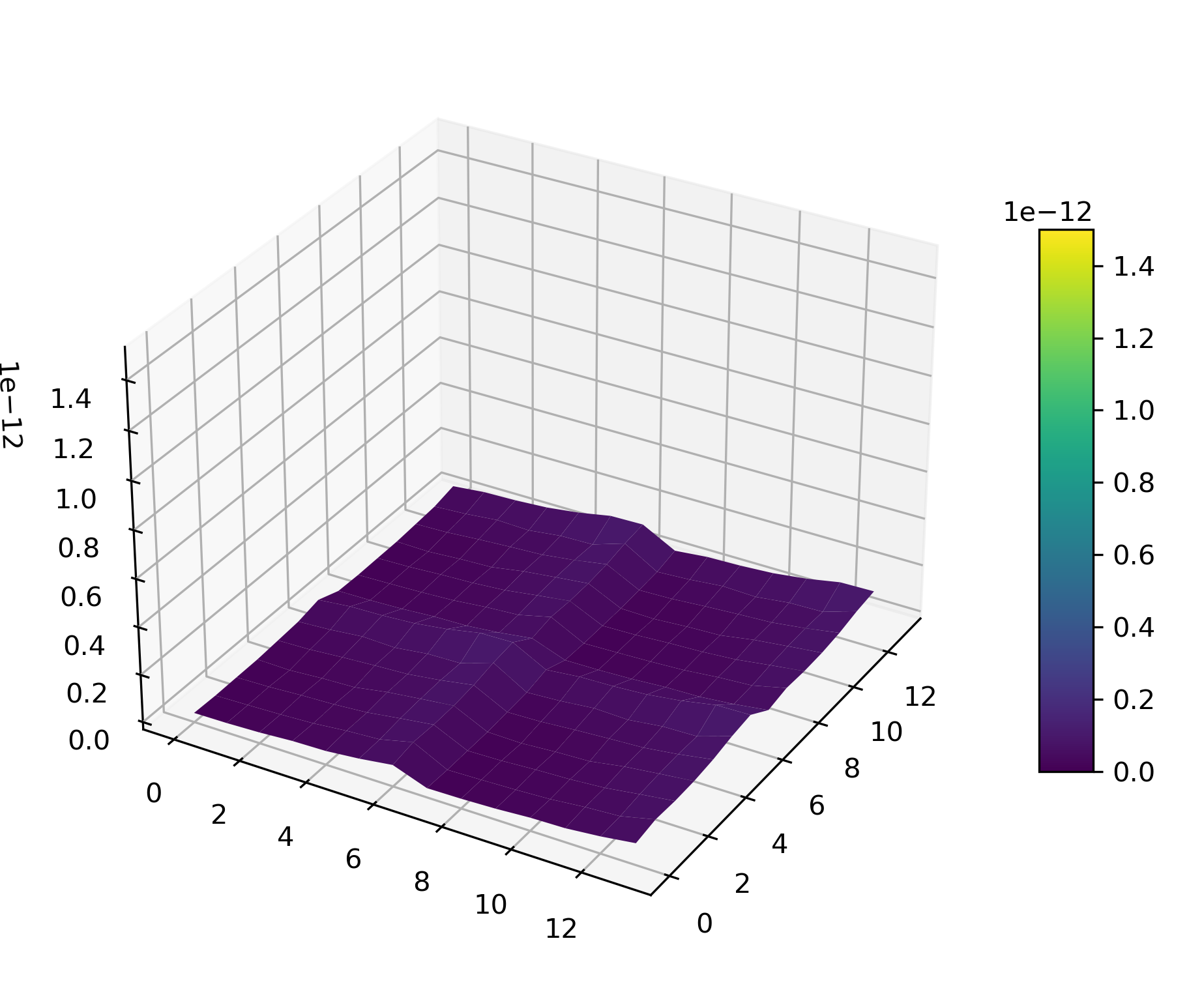}
		  \centerline{(b)$\gamma$=0.20}
	\end{minipage} \\
	%\qquad
	\begin{minipage}{0.49\linewidth}
		\centering
		\includegraphics[width=1\linewidth]{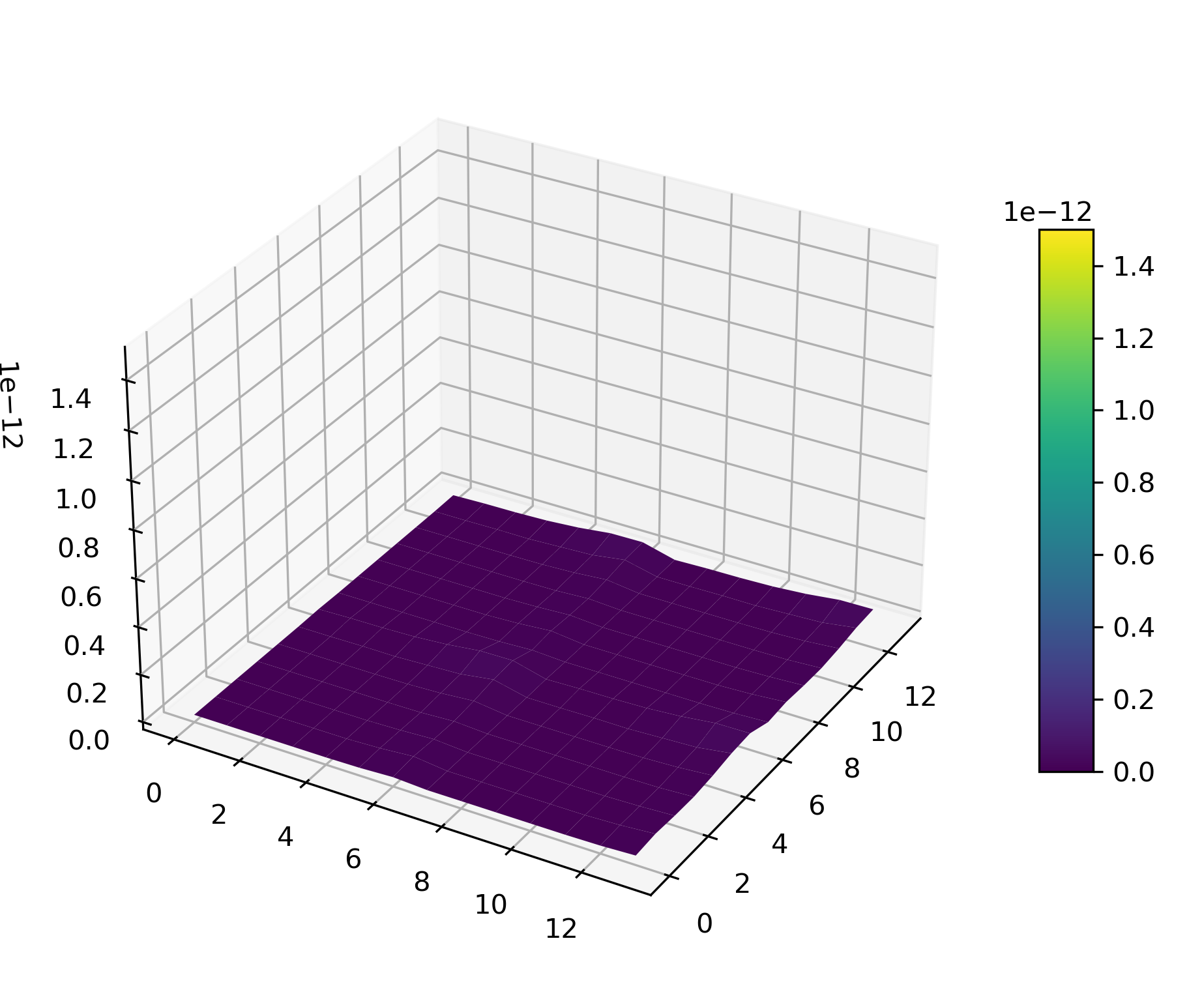}
		  \centerline{(c)$\gamma$=0.10}
	\end{minipage} 
	\begin{minipage}{0.49\linewidth}
		\centering
		\includegraphics[width=1\linewidth]{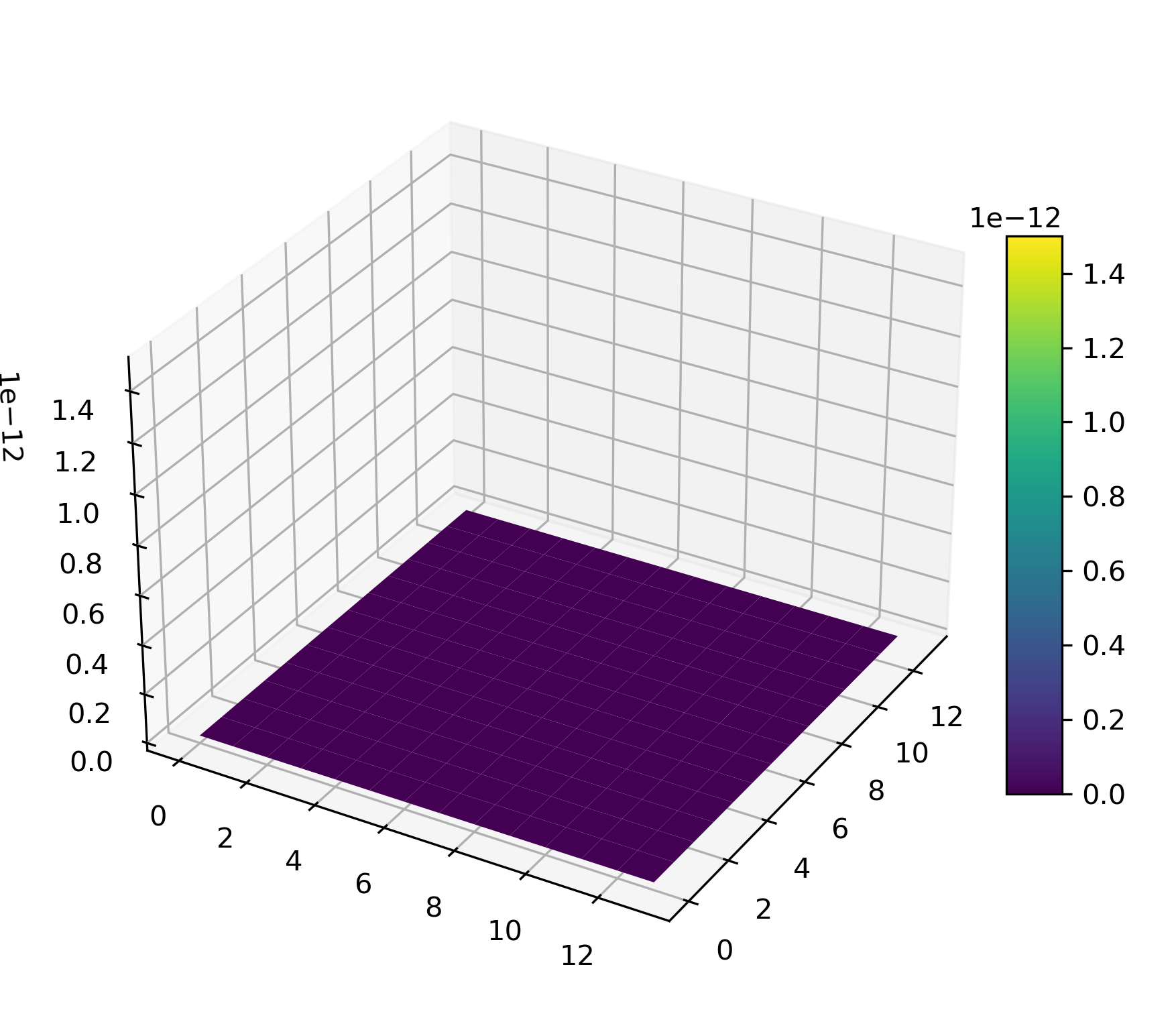}
		  \centerline{(d)$\gamma$=0}
	\end{minipage} 
\caption{VVI profiles of ROCKET-64 before and after time-domain optimization at $\gamma$ values of 0.20, 0.10 and 0.}
\label{fig:VVIprofile}
\end{figure}

To demonstrate the effectiveness of time-domain optimization, we stimulated severe switching currents with a correlation factor of 0.9 to model the operation of ROCKET-64. A total of 196 nodes from the ROCKET-64 chiplet layer, with each Rocket chiplet comprising 49 nodes, were selected as input for time-domain optimization. Due to the proximity to noise sources, on-chip decaps are more effective in mitigating SSN. Therefore, we focused on the effect of MOS capacitance changes on VVI, although on-interposer decaps can also be optimized in practice. Table~\ref{tab:time-domain} presents the results of time-domain optimization across various $\gamma$ values. A violation node is defined as one where the VVI exceeds zero. Initially, all nodes optimized solely in the frequency domain exhibited violations. As $\gamma$ decreases, reflecting a stricter tolerance for violations, achieving better optimization results required increasing the MOS capacitance, which led to a reduction in total VVI and number of violation nodes. Although the number of violation nodes did not decrease when $\gamma$ was set to 0.50 and 0.20, the VVI for each node was reduced, albeit not to zero, as illustrated in Fig.~\ref{fig:VVIprofile}. To achieve a total VVI of zero ($\gamma$=0), a minimum MOS capacitance of 17.8 nF was required. Due to the assumption of highly severe SSN conditions, a relatively large capacitance was necessary to eliminate voltage violations. However, under typical operating conditions, the required capacitance may be smaller. Therefore, in practical designs, designers can balance total allowable capacitance and VVI tolerance ($\gamma$) to meet the desired PDN performance. Based on this case study, without loss of generality, we have demonstrated the effectiveness of the proposed two-phase optimization method in achieving a robust PDN.

% \LLS{\section{Future Work}
% \label{future}
% Promising results have been demonstrated through comprehensive experimental evaluations of the proposed two-phase optimization method. However, there is potential for further improvements that could enhance its effectiveness and establish it as a significant area of research. Firstly, the compact modeling approach of the hierarchical PDN could benefit from cross-verification with more accurate full microwave analyses. This would help determine the optimal frequency range and limitations of the current modeling technique. Additionally, more realistic current models and correlation coefficients should be extracted from SPICE simulations of transmitter/receiver circuits under various workloads. Although the DNN model employed in this study already surpasses traditional DA and GA approaches, exploring more advanced DNN architectures could further improve performance and enhance transferability.
% }

\section{Conclusion}
\label{sec:conclusion}
In this paper, we propose an RL-based method to optimize the decap design of 2.5D hierarchical PDNs, integrating both frequency- and time-domain analyses. Through frequency-domain optimization, we successfully meet the target impedance requirements. Subsequent optimization using time-domain techniques notably mitigates SSN. By leveraging this two-phase optimization strategy, we significantly improve power integrity and achieve a robust PDN design. Experimental results highlight the importance of optimizing the decap capacitance and placement in 2.5D chiplet integration, demonstrating the efficacy of our proposed approach.


\end{document}